\documentclass[conference]{IEEEtran}
\IEEEoverridecommandlockouts
\usepackage{cite}
\usepackage{amsmath,amssymb,amsfonts}
\usepackage{algorithmic}
\usepackage{graphicx}
\usepackage{float}
\usepackage{textcomp}
\usepackage{xcolor}
\usepackage{fancyhdr}
\usepackage{subfigure}
\usepackage{url}
\usepackage{algorithm}
\usepackage{algorithmic}
\usepackage{hyperref}
\usepackage{verbatim}
\def\BibTeX{{\rm B\kern-.05em{\sc i\kern-.025em b}\kern-.08em
    T\kern-.1667em\lower.7ex\hbox{E}\kern-.125emX}}

\begin{document}
\pagestyle{fancy}

\fancyhead[C]{This paper appears on 2023 IEEE 97th Vehicular Technology Conference (VTC2023-Spring)}
\title{Mobile Edge Computing and AI Enabled Web3 Metaverse over 6G Wireless Communications: A Deep Reinforcement Learning Approach\vspace{-0.2cm}}

\begin{comment}
\author{\IEEEauthorblockN{Wenhan Yu}
\IEEEauthorblockA{Interdisciplinary Graduate Programme\\Nanyang Technological University\\
wenhan002@e.ntu.edu.sg }
\and
\IEEEauthorblockN{Xiaoming Shi}
\IEEEauthorblockA{Dalian University of Technology\\
shixm@mail.dlut.edu.cn}
\and
\IEEEauthorblockN{Terence Chua Jie}
\IEEEauthorblockA{Interdisciplinary Graduate Programme\\Nanyang Technological University\\
terencej001@e.ntu.edu.sg }
\and
\IEEEauthorblockN{Jun Zhao}
\IEEEauthorblockA{School of Computer Science \& Engineering\\ Nanyang Technological University\\
junzhao@ntu.edu.sg }
}
\end{comment}

\author{\IEEEauthorblockN{Wenhan Yu, Terence Jie Chua, Jun Zhao}\thanks{Wenhan Yu, Terence Jie Chua, Jun Zhao are with the School of Computer Science \& Engineering, Nanyang technological University, Singapore. Email: wenhan002@e.ntu.edu.sg, terencej001@e.ntu.edu.sg, JunZHAO@ntu.edu.sg}\vspace{-1cm}}

\maketitle
\thispagestyle{fancy}
\begin{abstract}
The Metaverse is gaining attention among academics as maturing technologies empower the promises and envisagements of a multi-purpose, integrated virtual environment. An interactive and immersive socialization experience between people is one of the promises of the Metaverse. In spite of the rapid advancements in current technologies, the computation required for a smooth, seamless and immersive socialization experience in the Metaverse is overbearing, and the accumulated user experience is essential to be considered. The computation burden calls for computation offloading, where the integration of virtual and physical world scenes is offloaded to an edge server. This paper introduces a novel Quality-of-Service (QoS) model for the accumulated experience in multi-user socialization on a multichannel wireless network. This QoS model utilizes deep reinforcement learning approaches to find the near-optimal channel resource allocation. Comprehensive experiments demonstrate that the adoption of the QoS model enhances the overall socialization experience.

\end{abstract}


\section{Introduction}

An immersive and interactive socialization experience on the Metaverse is fast becoming a reality, in which Extended Reality (XR) technologies enable users to interact with others through the mask of full-body avatars~\cite{social}. The usage of XR socialization is varied, and that includes many applications such as in education~\cite{teaching}, healthcare~\cite{medical}, for meetings~\cite{meetings} and many others. Web 3, also known as the decentralized web, is a new iteration of the internet that promises to be more secure, transparent, and open.

\begin{comment}
\begin{figure*}[t]
\centering
\includegraphics[width=.75\linewidth]{model.pdf}
\caption{System model.}
\label{fig:model}
\vspace{-0.4cm}
\end{figure*}
\end{comment}

\textbf{Motivation.}
To deliver a smooth socialization experience in Metaverse, physical and virtual worlds must be seamlessly integrated. Socialization involves accumulating experiences over time, so considering multiple users' experiences over an extended period is crucial. Users expect a certain level of performance and reliability from the applications and services they use. In Web3, service providers are required to provide service level agreements (SLAs) that specify the performance levels they will deliver to their users~\cite{web3}. Quality of service (QoS) is crucial to ensure that providers can meet these SLAs and maintain their reputation in the network. Decentralized applications (dApps) and services in Web3 demand low latency and high throughput, and given the decentralized structure of the Web3 network, a decentralized optimization method is necessary to consider the varying demands of different users. One approach is to use multiple servers for groups of users with different demands, and study each server to optimize its QoS to support dApps and services.

%Deep Reinforcement Learning (DRL) is effective for time-sequential problems, so we need a suitable Quality of Service (QoS) model for DRL to accommodate users' accumulated experience.

Virtual scenes on mobile devices are hindered by limited computing power. One solution is to use Mobile Edge Computing (MEC) to offload the integration task to an edge server. However, allocating channels to multiple users is a challenge. Competition between multiple users for limited channels often result in an inefficient allocation of channel resource. As such, our work seeks to tackle the user to channel allocation problem.

\textbf{Related work.} In recent years, the QoS over wireless networks have been thoroughly studied, including the metrics in cloud computing~\cite{QoScloud}, Internet of Things (IoT)~\cite{QoSiot}, and sensor networks~\cite{QoSsensor}. However, most of them consider modeling and optimizing the QoS of users in a single step, without considering the accumulated experience for socialization. In addition, there are existing works~\cite{du1, du2} that have proposed comprehensive optimization problems pertaining to the Metaverse and obtained elegant solutions. However, none of the above-mentioned works considered differing edge computing service requirements across users. In our work, we employed deep reinforcement learning (DRL) to solve our proposed problem as it is able to handle complex, sequential scenarios and achieve excellent performance~\cite{RLC}. 

\textbf{Approach.} We proposed a Quality of Service (QoS) based model to facilitate the socialization over wireless networks. Consider a sequential downlink scenario where a virtual service provider (VSP) generates different XR users' (XU) augmented scenes based on the composition of the physical and the Metaverse virtual environment, then transmit these scenes to the XUs across wireless network channels. Some XUs may not be arranged with any channel as there are limited channels. In this case, they can only generate graphically inferior augmented scenes locally, which will significantly impact the user experience. Based on the above-mentioned process, we devise a QoS model which accounts for relevant factors to optimize XU-VSP channel allocation. We adopt a DRL approach as the proposed problem is sequential, which involves multiple interactions between XUs and VSP, resulting in varying levels of XU satisfaction across the iterations.

\textbf{Why not convex optimization?} The problem (\ref{eq:M1}) is sequential and each XU's complaint count changes in each iteration. The changing complaint count across the rounds will influence the optimal solution. Therefore, model-based RL and convex optimization are not suitable methods to handle our proposed problem.

\textbf{Contributions.} Our contributions are as follows:

\begin{itemize}
\item \emph{Channel resource optimization with DRL methods:} We devised a novel Metaverse-based XR socialization scenario where DRL is used to guide XU-VSP allocation for XR scene computation over wireless networks.

\item \emph{Quality-of-Service model for accumulated user experience:} We introduced an innovative metric for the QoS model to balance multiple users' requirements to improve their accumulated experiences, and designed it to be easy-to-train for DRL methods.

\item \emph{DRL Scenario and Algorithm Design:} We crafted rational reinforcement learning state, action, environment, based on our proposed model, and employ multiple deep reinforcement learning approaches to find the optimal solution. Extensive experiments demonstrate that our proposed QoS model with DRL methods are reasonable to improve the user experience according to several metrics.
\end{itemize}

The rest of the paper is organized as follows. Section~\ref{models} introduces our system and quality-of-service models. Then Section~\ref{RL} proposes our deep reinforcement learning approach. In Section~\ref{experiment}, extensive experiments are performed, and various methods are compared to show the satisfactory performance of our strategy. Section~\ref{conclude} concludes the paper.

\section{System model}
\label{models}
Consider the indoor downlink transmission of a web 3 virtual service provider (VSP) providing $N$ XR device users (XU) $\mathcal{N}={1,2,...,N}$, with XR scenes which are generated based on the integration of physical and virtual world, via a set of $\mathcal{M} = {1,2,...,M}$ channels in a decentralized network. Each XU $n \in \mathcal{N}$ wants to transform its physical-world scene into a 3D virtual scene to interact with other users in it, and the web 3 technology allows for trustless and decentralized communication and processing. As the generation of XR scenes is computationally intensive, it will be optimal for users to offload the XR scene computation task to the VSP, instead of performing local scene computation. However, as there are limited VSP channels serving a much larger user (XU) base, some users may be arranged with no channel and have to generate scenes locally on their XR device. Evidently, ensuring real-time interactions while processing such intensive tasks locally will avianize the visual verisimilitude, ulteriorly decreasing the users' satisfaction. We next describe the communication and quality-of-service (QoS) models to illustrate the details of our system model in the context of web 3 technology.

\subsection{Communication Model}\label{sub:Communication-Model}
We first introduce the communication model based on a wireless network in a web 3 environment. The decentralized VSP will manage the allocation of its downlink channels $\mathcal{M}$ to all XUs $\mathcal{N}$ in a trustless and transparent manner using smart contracts on a blockchain. Furthermore, we denote $D^{t} = \{D_{1}^{t},D_{2}^{t},...,D_{N}^{t}\}$ as the data to be sent by VSP to XUs at time $t$. Specifically, $D_{n}^{t}$ is the data transmitted from VSP to XU $n$ at time ${t}$. The XU-VSP channel assignment is represented as a one-hot matrix
$
    \mathcal{L}^t=
    \begin{bmatrix}
    \ell_{1,0}^t & ... & \ell_{1,M}^t\\ 
    ... & ... & ...\\ 
    \ell_{N,0}^t & ... & \ell_{N,M}^t
    \end{bmatrix}
$
, and $\ell_{n,m}^t=\{0,1\}$  denotes whether XU $n$ is allocated to channel $m$. Specifically, $\ell_{n,0}^t=1$ means user $n$ is arranged with no channel and it needs to generate scenes locally. Given these, we can derive the downlink transmission delay as $d_{n}^{t}=\frac{D_{n}^{t}}{r_{n}^{t}}$, where the transmission rate is:
\begin{align}
&r_{n}^{t}=B\log_{2}\left(1+\frac{p_{n}h_{n,m}^{t}}{\sum\limits_{i\in\mathcal{N}\backslash\{n\},\ell_{i,m}^{t}=\ell_{n,m}^{t}=1}p_{i}h_{n,m}^{t}+B\sigma^{2}}\right).\label{eq:rate} 
\end{align}

$h_{n,m}^{t}=g_{n,m}^{t}l_n^{-\alpha}$ denotes the channel gain between XU $n$ and VSP in channel $m$, with $g_{n,m}^{t}$, $l_n$, $\alpha$ being the Rayleigh fading parameter, the distance between XU $n$ and VSP, and the path loss exponent, respectively. $p_{n}$ is the transmit power allocated by VSP for user $n$. $B$ is the bandwidth of each channel, and $\sigma^{2}$ denotes the one-sided noise power spectral density. It is evident that too many users sharing the same channel diminishes the effective transmission rate due to the acute interference, which may ultimately cause an unacceptable transmission delay. Therefore, making an effort to go in quest of an advisable management solution is of vital importance.

\subsection{Quality-of-Service (QoS) Model} \label{QoSmodel}
In the context of Web 3, users are multifarious and have different personalities, making their maximal tolerable delays (TDs) vary~\cite{QoS}. We assume that each user has a slightly different TD ($\mathcal{\tau}={\tau_{1},\tau_{2},...,\tau_{n}}$), and we define the satisfaction function $f(x)$ based on the delay exceedance coefficient represented by $\frac{d_{n}^{t}}{\tau_{n}}$. As transmission delay increases, the satisfaction function decreases continuously, albeit at a slowing rate. However, the satisfaction rate tapers off and will not be less than $-1$. We use $f(x)$ as the metric of the QoS.

In Web 3, where decentralized applications (dApps) run on blockchain networks, it is crucial to ensure a high QoS to attract and retain users. Therefore, we present our QoS model as:
\begin{align}
S_{n}^{t}(D_{n}^{t},p_{n},h_{n,m}^{t},\mathcal{L}^{t},\tau_{n})=\exp\left(-\frac{D_{n}^{t}}{r_{n}^{t}\tau_{n}}\right)-1.
\end{align}

where $S_{n}^{t}$ represents the satisfaction of user $n$ at time $t$, $D_{n}^{t}$ is the transmission delay, $p_{n}$ is the power consumed by user $n$, $h_{n,m}^{t}$ is the channel gain between user $n$ and VSP $m$ at time $t$, $\mathcal{L}^{t}$ is the channel allocation matrix at time $t$, and $\tau_{n}$ is the user's maximal tolerable delay.

The QoS model is used to determine the satisfaction of each user at each time step and to calculate the reward for each action taken by the RL agent. If $S_{n}^{t} \leq -0.5$, the user is marked as unsatisfied, and their dissatisfaction count increases. If they are assigned to local XR computation, which has worse graphics, frame rate, and is more energy-consuming, $S_{n}^{t}$ is set to $-0.5$. The maximum tolerable dissatisfaction count, $K_n$, is different for each user to simulate their different tolerances. If the dissatisfaction count $dissat_{n}^{t}$ reaches $K_n$, the user complains. The dissatisfaction count increases by 1 in each downlink transmission if the user is dissatisfied with the delay or service. Furthermore, we add a total complaint count constraint $\Gamma^t \leq Z$ to the system. When the total complaint counts of all users exceed the constraint, the RL agent will be given a huge penalty. We assign a large penalty for exceeding the complaint count constraint to simulate the impact of negative reputation on the provider of the dApp or service.

\subsection{Problem formulation}
To sum up, our goal is to find the optimal XU-VSP channel arrangement within $T$ steps, which maximizes the QoS under the delay and complaint constraints:
\begin{align}
\setlength{\belowdisplayskip}{4pt plus 1pt minus 1.0pt}
\setlength{\belowdisplayshortskip}{4pt plus 1pt minus 1.0pt}
\setlength{\abovedisplayskip}{4pt plus 1pt minus 1.0pt} \setlength{\abovedisplayshortskip}{0.0pt plus 2.0pt}
\max_{\boldsymbol{\mathcal{L}^{t}}}& \sum_{t=1}^{T} \sum_{n\in\mathcal{N}}S_{n}^{t}(D_{n}^{t},p_{n},h_{n,m}^{t},\mathcal{L}^{t},\tau_{n})\label{eq:M1}\\
s.t. ~~& \sum_{n\in \mathcal{N}} \lfloor \frac{dissat_{n}^{T}}{K_n} \rfloor \leq Z, \label{eq:st1} \\
& \sum_{j=1}^{M}\ell_{n,j}^{t}\leq 1, \forall n \in \mathcal{N}, \forall t \in \mathcal{T}, \label{eq:st2}\\
& \sum_{i=1}^{N}\ell_{i,m}^{t}>0, \forall m \in \mathcal{M}, \forall t \in \mathcal{T}. \label{eq:st3}
\end{align}

Constraint (\ref{eq:st1}) is the total complaint constraint. This means XU $n$ will complain after being dissatisfied for $K_n$ times, and the total number of complaint times of all XUs should be less than $Z$. Constraint (\ref{eq:st2}) restrics each XU to only be allocated to a maximum of one channel. Moreover, not every action in each state is equally advisable, so we set a constraint (\ref{eq:st3}) to eliminate actions that represent channels not occupied by any XU.

% we do not know when the tolerant complaint times will run out. 

\section{Deep reinforcement learning approach}
\label{RL}
\subsection{Reinforcement learning approach of our model}\label{RLapproach}
The ingenious design of the state, action spaces, and reward function is the key to the successful adoption of reinforcement learning methods. Now, we will expound on our approach from these three parts.
\subsubsection{State}
In reinforcement learning, attributes which are changing over time influences agents significantly. Therefore, (1) the data to be transmitted in each time step and (2) the dissatisfaction counts of each user in each time step, must be part of the state. In addition, as (3) transmission power and (4) channel gain greatly influence the transmission rates, they are recommended to be added. Then, we finally fix our state to contain  these four attributes:
\begin{align}
    &S^{t} = \{D_{n}^{t},dissat_{n}^{t},p_{n},h_{n,m}^{t}\},
    ~(n\in \mathcal{N}, t\in \mathcal{T}).\label{state}
\end{align}

\subsubsection{Action}
In our proposed scenario, the action taken is the channel arrangement selected by the VSP, which is:
\begin{align}
    &A^{t}=\mathcal{L}^{t}=
    \begin{bmatrix}
    \ell_{1,0}^t & \ell_{1,1} & ... & \ell_{1,M}^t\\ 
    \ell_{2,0}^t & \ell_{2,1} & ... & \ell_{2,M}^t\\
    ... & ... & ... & ...\\ 
    \ell_{N,0}^t & \ell_{N,1} & ... & \ell_{N,M}^t
    \end{bmatrix},
    (t\in\mathcal{T}).
    \label{action}
\end{align}
Each $\ell_{n,m}^t(\forall n\in\mathcal{N},\forall m\in\mathcal{M}\cup\{0\})$ is a one-hot indicator whose value is $0$ or $1$. In discrete RL, we need to encode the actions into serial discrete numbers. However, the number of discrete actions should be $2^{N\times (M+1)}$, which is tremendous. Therefore, we re-design the element of actions as $\hat{\ell}_{n}^t \in \{0,1,...,M\}~(\forall n\in \mathcal{N})$, and encode them into discrete action indexes. This reduces the dimension of actions from $2^{N\times(M+1)}$ into $(M+1)^N$. 

\begin{comment}
\begin{figure}[t]
\centering
\includegraphics[width=0.75\linewidth]{action.pdf}
\vspace{-0.3cm}
\caption{Discrete action encoding.}
\label{fig:action}
\vspace{-0.4cm}
\end{figure}
\end{comment}

\subsubsection{Reward}
Environments with sparse rewards typically impedes training progress. To facilitate training, we provide the agent with more consistent feedbacks and intricate rewards. Thus, (1) we give the average satisfaction of each XU as a part of the reward, and (2) a penalty every time a complaint occurs. Moreover, (3) a small penalty is given for the wastage of channel resources (if there exist a channel with no user in any iteration). (4) If the total complaint counts exceeds the acceptable limit, a large penalty will be issued to the agent and the episode will be terminated early. In summary, our reward function is as follows:\\
\begin{equation}
R^t= \label{eq:reward}
\begin{cases}
\Bar{S_n^t}, & \text{ if } \ell_{n,0}^t=0, n\in \mathcal{N}, t\in \mathcal{T},\\
 -0.5,& \text{ if } \ell_{n,0}^t=1, n\in \mathcal{N}, t\in \mathcal{T},\\
 -2,& \text{ if } dissat_n^t=K,\\
 -(T-t),& \text{ if } \Gamma^t=Z.
\end{cases}
\end{equation}

\subsection{Algorithms} \label{algorithms}
\subsubsection{PPO-based network}
Proximal Policy Optimization (PPO) by openAI \cite{PPO} is an enhancement of the traditional policy gradient algorithm. In dealing with sequential problems like reinforcement learning, slight changes in parameters can influence performance sigificantly and it is difficult to tune the parameters to achieve a satisfactory performance. PPO counters the problem of highly sensitive and noisy advantage estimates by taking a conservative approach, using the Kullback–Leibler (KL) divergence penalty to constrain the policy change. PPO also takes advantage of an importance sampling strategy~\cite{kahn1951estimation}, using asynchronous policies for training and sampling data to improve sampling efficiency. The loss function of the Actor is defined~\cite{PPO} as:
\begin{align}
    L^{CLIP}(\theta)=E_{t}[\min(r_{\theta}A_{t},clip(r_{t}(\theta),1-\epsilon,1+\epsilon)A_{t})].\nonumber
\end{align}
\begin{itemize}
    \item $\theta$: The policy
    \item ${E_{t}}$: The empirical expectation over time
    \item $r_{t}$: The ratio of current to old policy
    \item $A_{t}$: The advantage estimation at $t$
\end{itemize} This clipping approach prohibits the huge bias and keeps the policy within a trust region. %The PPO algorithm is shown in Algorithm~\ref{alg:PPO}.

\subsubsection{Baselines}
We implement commonly adopted reinforcement learning algorithms that can handle discrete action space.

\begin{itemize}
    \item \textit{DQN. } DQN~\cite{RL1} estimates the $Q$-values for each discrete action by a neural network. It also uses a replay buffer to store trajectories, and is trained with random batches of sampled data.
    \item \textit{Dueling-DQN}. Dueling DQN \cite{duelDQN} is a notable DQN-variant that uses two streams to separately estimate the scalar state value ($V$) and the advantages ($A$) for each action, and utilizes these values to produce a final output value.
    \item \textit{A3C. }A3C is a distributed advantage actor-critic network which updates asynchronously. In A3C\cite{A3C}, multiple workers sample data and train their policies within their environments. Meanwhile, with the help of the advantage actor-critic structure (which uses the value function as the baseline and the exceeding part as the advantage), it has lower variance and significantly improves robustness.
    \item \textit{A2C. }A2C is a synchronized version of A3C. According to the work by OpenAI~\cite{A2C}, A2C performs better than A3C in some tasks as it ensures the global policy is always the optimal one. Furthermore, the output by A2C is more stable and improves the convergence rate.
\end{itemize}
\vspace{-0.2cm}

\subsection{Metrics}
We illustrate the performance of our proposed methods in tackling our proposed problem through comparing our algorithms against several key metrics.
\begin{itemize}
    \item \textit{Total Reward}. The RL reward designed in Section~\ref{RL} describes the metric clearly. The reward is an overarching measure of the performance of our proposed models.
    \item \textit{Average Rate}. The average rate of XUs is the average downlink transmission rate of XUs which are allocated with channels. It is an important metric as a higher average rate directly implies better XU-VSP channel allocation.
    \item \textit{Channel Resource Waste Counts}. The channel resource waste counts (total number of occasions where there are channels allocated with no users) reflect the channel utilization. A high channel resource waste count reflects poor allocation of channel resources.
    \item \textit{Successful transmission counts}. The successful transmission counts shows how many downlink transmissions can be completed successfully before breaking the complaint constraint within $T_{max}$ steps. A higher successful transmission count indicates an overall better handling of clients' request.
\end{itemize}
\vspace{-0.2cm}

\begin{comment}
\begin{figure}[!t]
        \renewcommand{\algorithmicrequire}{\textbf{Input:}}
        \renewcommand{\algorithmicensure}{\textbf{Output:}}
        \begin{algorithm}[H]
            \caption{\label{alg:PPO}Proximal Policy Optimization}
            \begin{algorithmic}[1]
                \REQUIRE policy parameters $\theta_0$, value function parameters $\phi_0$
                \FOR{$i = 0,1,2...$}
                    \STATE Collect trajectories $D_i={\tau_k}$ by running policy $\pi_i=\pi(\theta_i)$.
                    \STATE Calculate reward $R_t$ based on the reward function.
                    \STATE Compute advantage estimate $A_t$ using Generalized Advantage Estimation (GAE) method based on current value function $V_{\phi_i}$.
                    \STATE Update policy by maximizing PPO-Clip objective with Adam:\\
                    \begin{center}
                    $\theta_{i+1}=\mathop{\arg\max}\limits_{\theta} \frac{1}{|D_i|T} \sum\limits_{\tau \in D_i} \sum\limits_{t=0}^T$
                    $\min(\frac{\pi_\theta(a_t|s_t)}{\pi_{\theta_i}(a_t|s_t)} A^{\pi_{\theta_i}}(s_t,a_t),g(\epsilon, A^{\pi_{\theta_i}}(s_t,a_t)))$
                    \end{center}
                    \STATE Fit value function by regression on mean-squared error via gradient descent method:\\
                    $\phi_{i+1}=\mathop{\arg\min}\limits_\phi \frac{1}{|D_i|T} \sum\limits_{\tau \in D_i} \sum\limits_{t=0}^T (V_\phi(s_t)-R_t)^2$
                \ENDFOR
            \end{algorithmic}
        \end{algorithm}
\vspace{-0.8cm}
\end{figure}
\end{comment}

\section{Experiment results}
\label{experiment}
\subsection{Configuration} \label{Configuration}
In order to evaluate the QoS model's ability to handle settings with varying number of clients, we fix the number of channels to $3$ and conduct experiments with the number of XUs ranging from $4$ to $7$. The bandwidth and noise are simulated to be $B=5$ MHz and $\sigma^2=-100$ dBm, respectively. We initialize different tolerable delay limit values, transmission powers for each XU at random, from a uniform distribution of $(30, 70)$ ms and $(0.5,2.0)$ Watt, respectively. The channel gain $g_{n,m}^t$ follows the Rayleigh distribution and $\alpha=2$ is the pass loss exponent. The XR scene data sizes $D_n^t$ to be downloaded by each XUs, vary uniformly across $10$ to $30$ Mb for each episode. Each users' maximum tolerable complaint count is initialized with a random value from a uniform distribution of $(5,10)$. The system's total tolerable complaint count limit is set to be $Z=5$. We set the maximum time step in each episode to be $100$. In order to better compare different algorithms, all experiments are conducted under the same random seed. We train the models for 40000 episodes, and each episode encompasses an entire execution of the optimization problem. The episode ends when the maximum time steps per episode or when the system's total tolerable complaint count limit, has been reached. And for A3C, the number of multi-threads is set to be 3.

% At the start of each time step, we set the satisfaction level $S_{def}$ of each XU to a default value of $-0.5$, which is equivalent to the satisfaction level of XUs arranged with no channel.

\subsection{Result analysis}
\begin{figure*}[t]
\centering
\subfigtopskip=2pt
\subfigbottomskip=2pt
\subfigure[Average rate under different scenarios.]{
\begin{minipage}[t]{0.3\linewidth}
\centering
\includegraphics[width=1\linewidth]{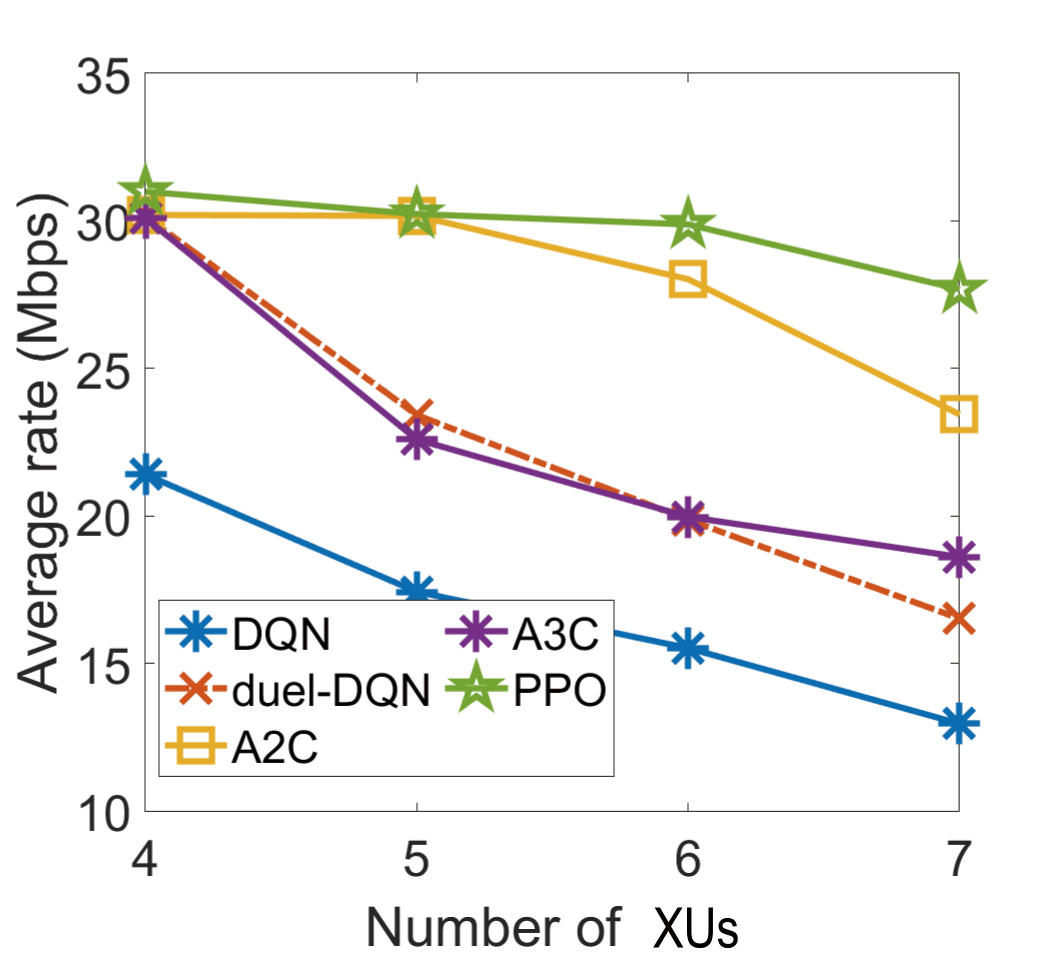}
\label{fig:rate}
\vspace{-10mm}
\end{minipage}
}%
\subfigure[Channel waste times with user number.]{
\begin{minipage}[t]{0.3\linewidth}
\centering
\includegraphics[width=1\linewidth]{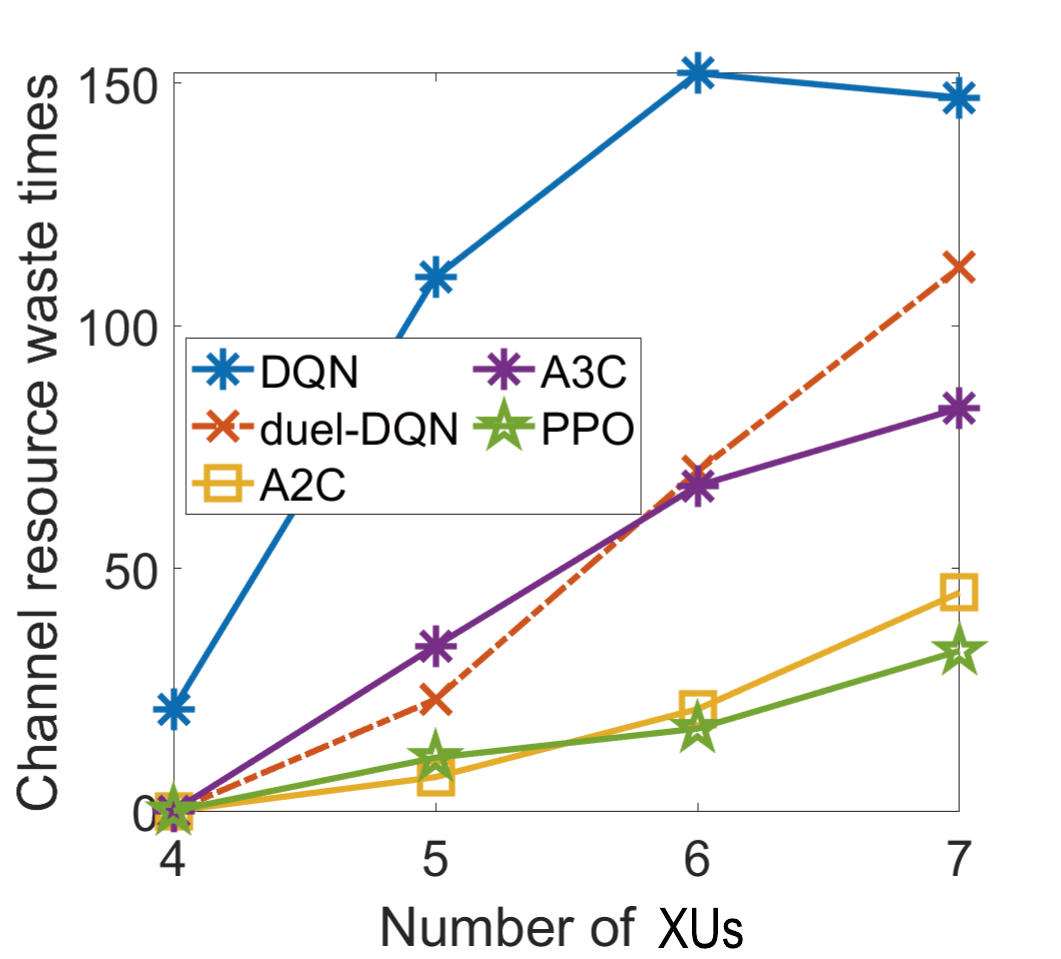}
\label{fig:waste}
\vspace{-10mm}
\end{minipage}%
}%
\subfigure[Successful steps with user number.]{
\begin{minipage}[t]{0.3\linewidth}
\centering
\includegraphics[width=1\linewidth]{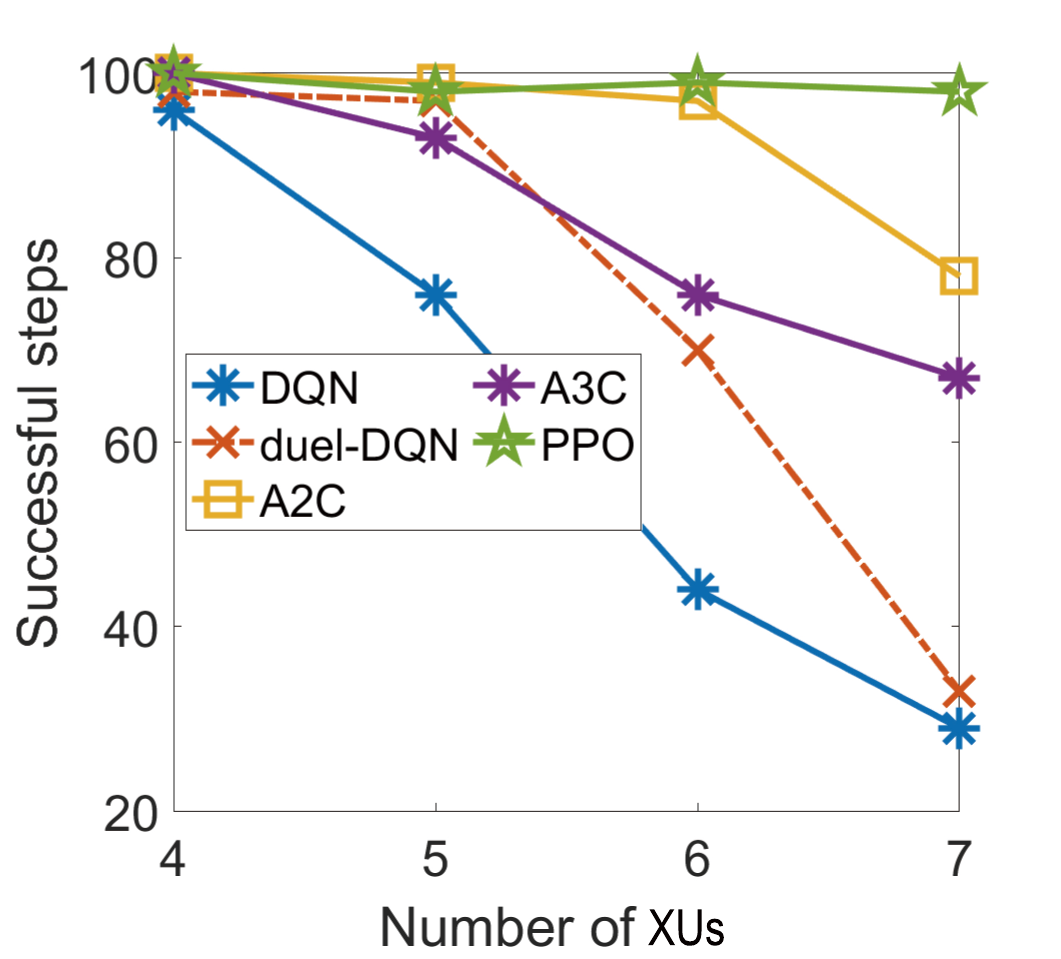}
\label{fig:step}
\vspace{-10mm}
\end{minipage}%
}%
\caption{Metrics under different scenarios and algorithms.}
\label{fig:metric}
\vspace{-0.5cm}
\end{figure*}

\begin{figure}[t]
\centering
\subfigtopskip=2pt
\subfigbottomskip=2pt
\subfigure[The average rewards.]{
\begin{minipage}[t]{0.49\linewidth}
\centering
\includegraphics[width=1\linewidth]{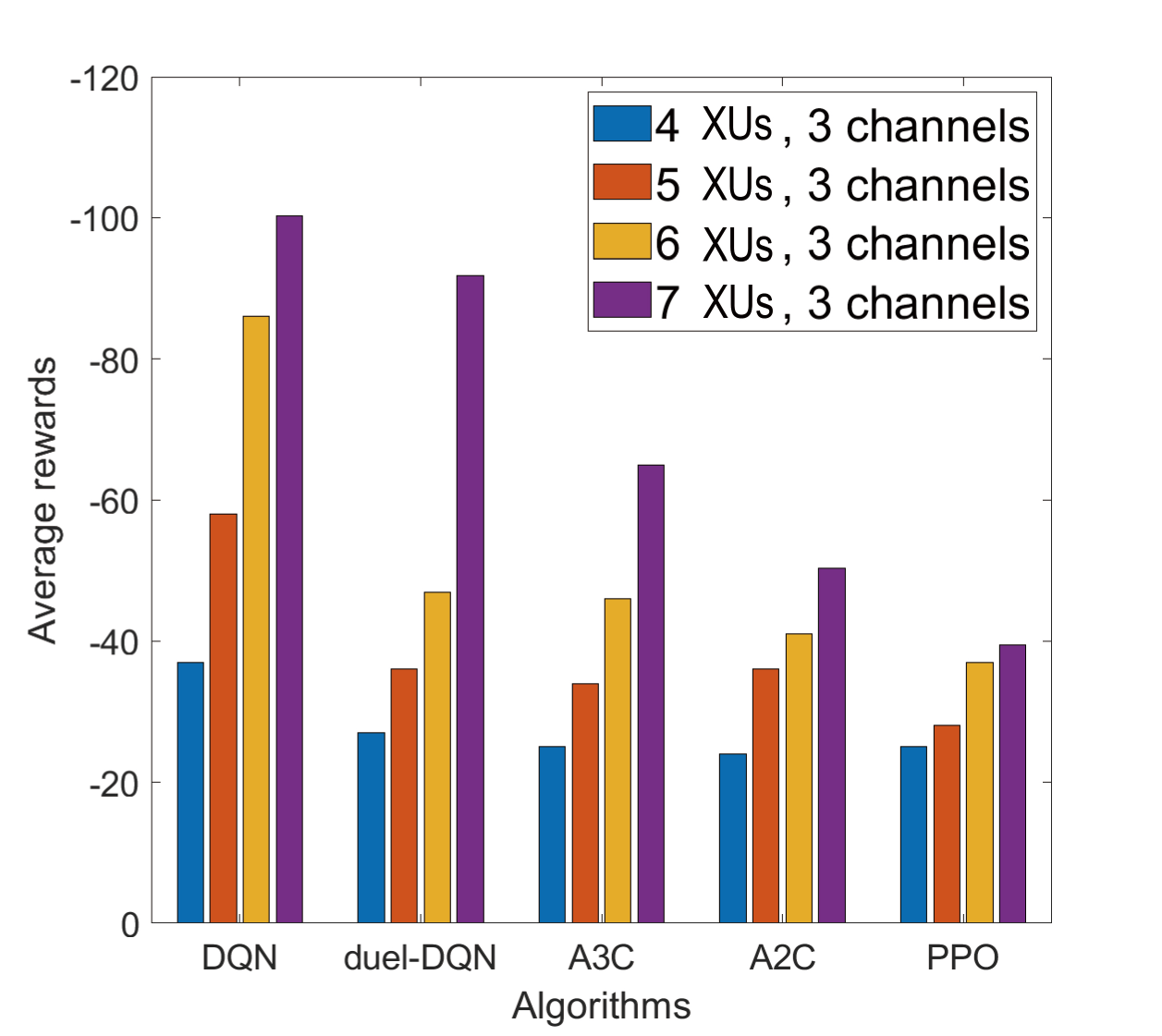}
\label{fig:ave}
\vspace{-10mm}
\end{minipage}
}%
\subfigure[Reward during training of 7 XUs, 3 channels.]{
\begin{minipage}[t]{0.49\linewidth}
\centering
\includegraphics[width=1\linewidth]{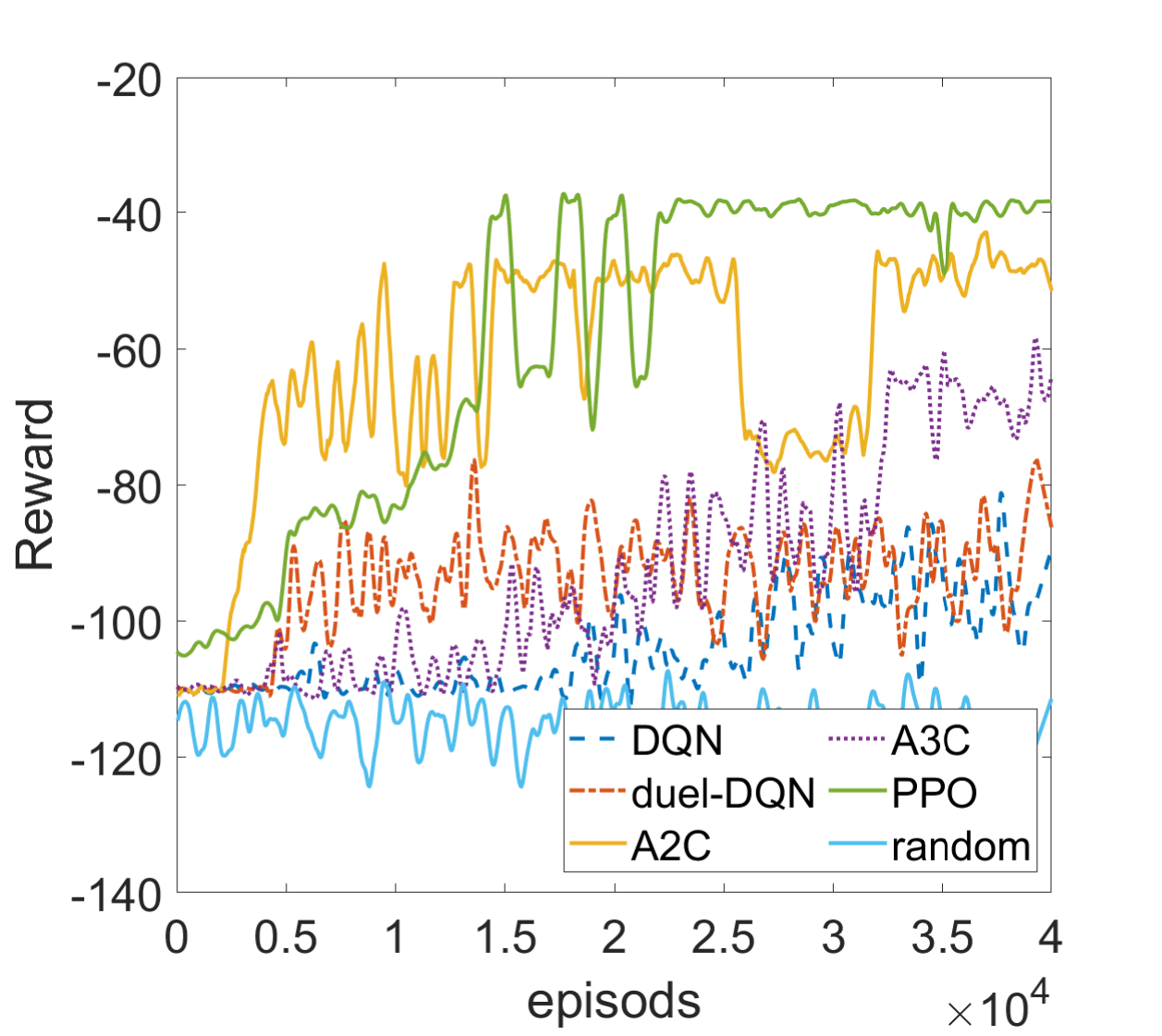}
\label{fig:37}
\vspace{-10mm}
\end{minipage}%
}%

\caption{The RL reward during training.}
\label{fig:reward}
\vspace{-0.5cm}
\end{figure}

Four algorithms (DQN, duel-DQN, A3C, A2C) and a random policy under different scenarios are compared to show the viability of our QoS model. We present the performances of the different algorithms based on reward obtained (shown in Fig.~\ref{fig:reward}) and other key metrics (Fig.~\ref{fig:metric}).

% We illustrate the figures about RL reward during training in Fig.~\ref{fig:reward}, and the metrics under different scenarios and algorithms in Fig.~\ref{fig:metric}.

Fig.~\ref{fig:ave} shows the average rewards across the final 2000 epochs, achieved by each algorithm, which is the indicative performance of each algorithm. The rewards obtained by the algorithms are similar in scenarios with fewer users, but they vary a lot in more complex scenarios. Although, DQN and duel-DQN perform the worst among the five algorithms, the addition of the dueling component to a traditional DQN still provides a boost in performance. This is reflected in the comparatively higher reward obtained by the duel-DQN in all scenarios, when compared to the traditional DQN. However, in scenarios with more than 5 XUs, they both fail to find a good solution. 

Both policy gradient algorithms (PPO and A2C) perform much better than the DQN-based algorithms. Although A3C uses multiple workers to improve sample efficiency, the performance of the algorithm is not stable. This is reflected in the poorer performance of A3C when compared to A2C and PPO. Another plausible reason behind a better performing A2C is that A2C utilizes Graphic Processing Units (GPUs) more effectively through batch updates~\cite{A2C}.

% A3C performs worse when compared to A2C and PPO as Compared to A2C and PPO, A3C is worse at handling high exploration complexity. Although it uses multiple processes to improve the sample efficiency, it is not so stable in our scenario. We assume the reason that A2C works better than A3C in our problem is that A2C can more effectively use GPUs, and it is more cost-effective when using a single-GPU machine~\cite{A2C}.

The rewards obtained by each algorithm (setting with $7$ XU) in each episode, during the training process is illustrated in Fig.~\ref{fig:37}. It shows the convergence speed and stability of different algorithms. Although the dueling structure in duel-DQN is expected to accelerate the convergence speed, both DQN and the duel-DQN achieve low final rewards reflecting poor model convergence. This poor performance obtained by the DQN-based algorithms could be attributed to the extreme difficulty of predicting the value of all actions in a huge action space scenario. When compared to PPO, A2C has a greater risk of falling into the local minimum, which demonstrates that the policy change constraint in PPO makes it more stable in our proposed scenarios.

% Since DQN-based algorithms need to evaluate all actions under states, this could be attributed to the extreme difficulty of predicting the value of all actions in such a huge action space scenario.

When comparing the algorithms against important metrics, we note that PPO is the superior algorithm. PPO achieves more optimal solutions when compared to the other algorithms. Furthermore, the prowess of PPO and A2C is further accentuated in their ability to decrease channel resource waste counts, despite channel waste counts being a small part within the overall objective (shown in Eq.(\ref{eq:reward})). In contrast to PPO and A2C, the other three algorithms are unable to learn effective strategies as they are unable to handle the large action space.

% In regard to the metrics, Fig.~\ref{fig:rate},~\ref{fig:waste},~\ref{fig:step} show the average transmission rates of XUs, the channel waste times, and the successful steps, respectively. They demonstrate the prowess of PPO in optimizing our problems. Among them, the channel resource waste times is a relatively small part of our reward function Eq.~(\ref{eq:reward}), however, PPO and A2C can still give heed to it and decrease the waste times. 

\section{conclusion}
\label{conclude}
 We present an XR socialization over wireless network problem, and proposed a novel QoS model to tackle it. Traditional optimization strategy are not suitable to tackle the proposed sequential problem, so deep reinforcement learning approaches are adopted. Multiple reinforcement learning algorithms are compared, and the experiments demonstrate that PPO performance the best in our scenario.

 \section*{Acknowledgement}

This research is partly supported by the Singapore Ministry of Education Academic Research Fund under Grant Tier 1 RG90/22, Grant Tier 1 RG97/20, Grant Tier 1 RG24/20 and Grant Tier 2 MOE2019-T2-1-176; and partly by the NTU-Wallenberg AI, Autonomous Systems and Software Program (WASP) Joint Project.

{\small
\bibliographystyle{IEEEtran}
%\bibliography{ref}
% Generated by IEEEtran.bst, version: 1.14 (2015/08/26)

}

\end{document}